\documentclass{article}

\usepackage{PRIMEarxiv}
\usepackage{fontawesome5}  

\pdfoutput=1

\usepackage[utf8]{inputenc} 
\usepackage[T1]{fontenc}    
\usepackage{hyperref}       
\usepackage{url}            
\usepackage{booktabs}       
\usepackage{amsfonts}       
\usepackage{nicefrac}       
\usepackage{microtype}      
\usepackage{lipsum}
\usepackage{fancyhdr}       
\usepackage{graphicx}       
\graphicspath{{media/}}     
\usepackage{setspace}
\usepackage{float}
\usepackage{multirow}
\usepackage{amsmath}

\title{\textbf{DDCSR: A Novel End-to-End Deep Learning Framework for Cortical Surface Reconstruction from Diffusion MRI}

\author{
  Chengjin Li  \thanks{University of Electronic Science and Technology of China, Chengdu, China.} \\
  \texttt{chengjinli@std.uestc.edu.cn} \\
  \And
  Yuqian Chen\thanks{Brigham and Women’s Hospital, Harvard Medical School, Boston, USA.} \\
  \texttt{ychen128@bwh.harvard.edu} \\
  \And
  Nir A. Sochen\thanks{School of Mathematical Sciences, University of Tel Aviv, Tel Aviv, Israel.} \\
  \texttt{sochen@tauex.tau.ac.il} \\
  \And
  Wei Zhang\footnotemark[1] \\
  \texttt{zhangwei@uestc.edu.cn} \\
  \And
  Carl-Fredrik Westin\footnotemark[2] \\
  \texttt{WESTIN@bwh.harvard.edu} \\
  \And
  Rathi Yogesh\footnotemark[2] \\
  \texttt{YOGESH@bwh.harvard.edu} \\
  \And
  Lauren J O’Donnell\footnotemark[2] \\
  \texttt{odonnell@bwh.harvard.edu} \\
  \And
  Ofer Pasternak\footnotemark[2] \\
  \texttt{ofer@bwh.harvard.edu} \\
  \And
  Fan Zhang  \footnotemark[1]~~ \textsuperscript{\faEnvelope} \\
  \texttt{fan.zhang@uestc.edu.cn} \\
}

}

\begin{document}
\maketitle

\begin{abstract}
Diffusion MRI (dMRI) plays a crucial role in studying brain white matter connectivity. Cortical surface reconstruction (CSR), including the inner whiter matter (WM) and outer pial surfaces, is one of the key tasks in dMRI analyses such as fiber tractography and multimodal MRI analysis. Existing CSR methods rely on anatomical T1-weighted data and map them into the dMRI space through inter-modality registration. However, due to the low resolution and image distortions of dMRI data, inter-modality registration faces significant challenges. This work proposes a novel end-to-end learning framework, \textit{DDCSR}, which for the first time enables CSR directly from dMRI data. DDCSR consists of two major components, including: (1) an implicit learning module to predict a voxel-wise intermediate surface representation, and (2) an explicit learning module to predict the 3D mesh surfaces. Compared to several baseline and advanced CSR methods, we show that the proposed DDCSR can largely increase both accuracy and efficiency. Furthermore, we demonstrate a high generalization ability of DDCSR to data from different sources, despite the differences in dMRI acquisitions and populations.
\end{abstract}

\keywords{Diffusion MRI   \and Cortical Surface Reconstruction  \and Deep Learning \and Weak Supervision}

\section{Introduction}
Cortical surface reconstruction (CSR) of the inner white matter (WM) and the outer pial surfaces is a key task in neuroimaging analysis. Diffusion MRI (dMRI) is an advanced imaging technique that enables the quantification of tissue microstructural features \cite{pierpaoli1996toward} and \textit{in vivo} mapping of WM fiber tracts \cite{zhang2022quantitative}. In dMRI, CSR is a crucial step for many computational applications such as WM tractography \cite{shastin2022surface} and multimodal MRI analysis \cite{yeo2011organization}. However, the acquisition protocols of dMRI data, designed specifically to capture water molecule motion in brain tissues, inevitably lead to local distortions, noise artifacts, and reduced spatial resolution \cite{wu2008comparison,malinsky2013registration,jones2010twenty,albi2018image}. As a result, accurate and efficient CSR in dMRI continues to be a challenging task to be resolved.

Currently, the widely used dMRI CSR approaches are based on anatomical T1-weighted (T1w) data, from which surface reconstruction is performed using tools like FreeSurfer \cite{fischl2012freesurfer} and then non-linearly registered to the dMRI space. However, these approaches are time-consuming and highly rely on the quality of inter-modality registration, which is affected by local distortions \cite{wu2008comparison,jones2010twenty,albi2018image} and the lower resolution of dMRI data \cite{malinsky2013registration}. Furthermore, these approaches are not applicable when anatomical T1w data is unavailable. Thus, there is a growing demand for CSR methods that can operate directly on dMRI data. Although no such methods exist yet, two areas of research described below show significant potential to achieve this goal.

First, recent advances in dMRI have demonstrated the ability to successfully perform tasks like tissue segmentation and cortical parcellation directly from dMRI data \cite{ciritsis2018automated,theaud2022doris,yap2015brain,zhang2021deep,zhang2023ddparcel}, tasks that previously relied heavily on anatomical T1w data. A common feature of these studies is the utilization of the unique multi-parameter characteristics of dMRI, where each parameter (e.g., fractional anisotropy; FA, and mean diffusivity; MD) can offer greater discriminative power in describing specific brain structures. These multi-parameter characteristics hold significant potential for CSR directly on dMRI data, as the process fundamentally relies on accurately defining the boundaries between different tissue types. 

Second, several recent works have shown the promise of using deep learning to enable fast and accurate T1w-based CSR \cite{bongratz2022vox2cortex,bongratz2024neural,cruz2021deepcsr,henschel2020fastsurfer,hoopes2022topofit,lebrat2021corticalflow,ma2022cortexode,ma2023conditional,ma2024weakly,ma2021pialnn,wang2023ibeat,yalcin2024surfflow}. These methods can be divided into two categories, i.e., \textit{implicit and explicit}. The implicit methods \cite{cruz2021deepcsr,gopinath2023cortical,henschel2020fastsurfer,wang2023ibeat} predict intermediate voxel-wise surface representations (e.g., signed distance function, SDF), followed by post-processing methods such as topology correction using fast marching algorithm \cite{bazin2007topology} and mesh extraction using marching cubes (MC) \cite{lorensen1998marching}. While the implicit methods can improve the efficiency compared to traditional methods (e.g. FreeSurfer), the performance is highly affected by the partial volume effect in low-resolution data. On the other hand, the explicit methods directly generate 3D surfaces by optimizing mesh deformation. The majority of existing methods are supervised learning approaches that rely on ground truth (GT) surfaces generated using traditional methods \cite{bongratz2022vox2cortex,bongratz2024neural,hoopes2022topofit,lebrat2021corticalflow,ma2022cortexode,ma2023conditional,ma2021pialnn,yalcin2024surfflow}. One recently proposed CoSeg method \cite{ma2024weakly} provides a novel weakly supervised learning framework that uses voxel-level segmentation results as pseudo GT (pGT), without requiring precise GT surfaces. This indicates that combining implicit and explicit learning offers sufficient information for CSR, presenting a viable solution for dMRI-based CSR, which faces challenges including low resolution and the absence of GT surfaces.

This paper proposes a novel end-to-end deep learning framework to perform dMRI-based CSR, namely \textit{DDCSR}. This paper has three main contributions. First, we believe that this is the first deep-learning approach that enables automated CSR from dMRI without the need for additional anatomical MRI data. Second, we design a novel end-to-end framework that includes an implicit surface prediction network that learns SDFs of the WM and pial surfaces from multiple dMRI-derived feature maps, and an explicit surface reconstruction network that learns 3D mesh surfaces from the voxel-wise SDF. Third, we demonstrate a high generalization to data from different sources. The codes for DDCSR are released publicly at \href{https://github.com/ChengjinLii/DDCSR}{https://github.com/ChengjinLii/DDCSR}.

 \section{\textbf{Method} }

\begin{figure}
\centering
\includegraphics[width=\textwidth]{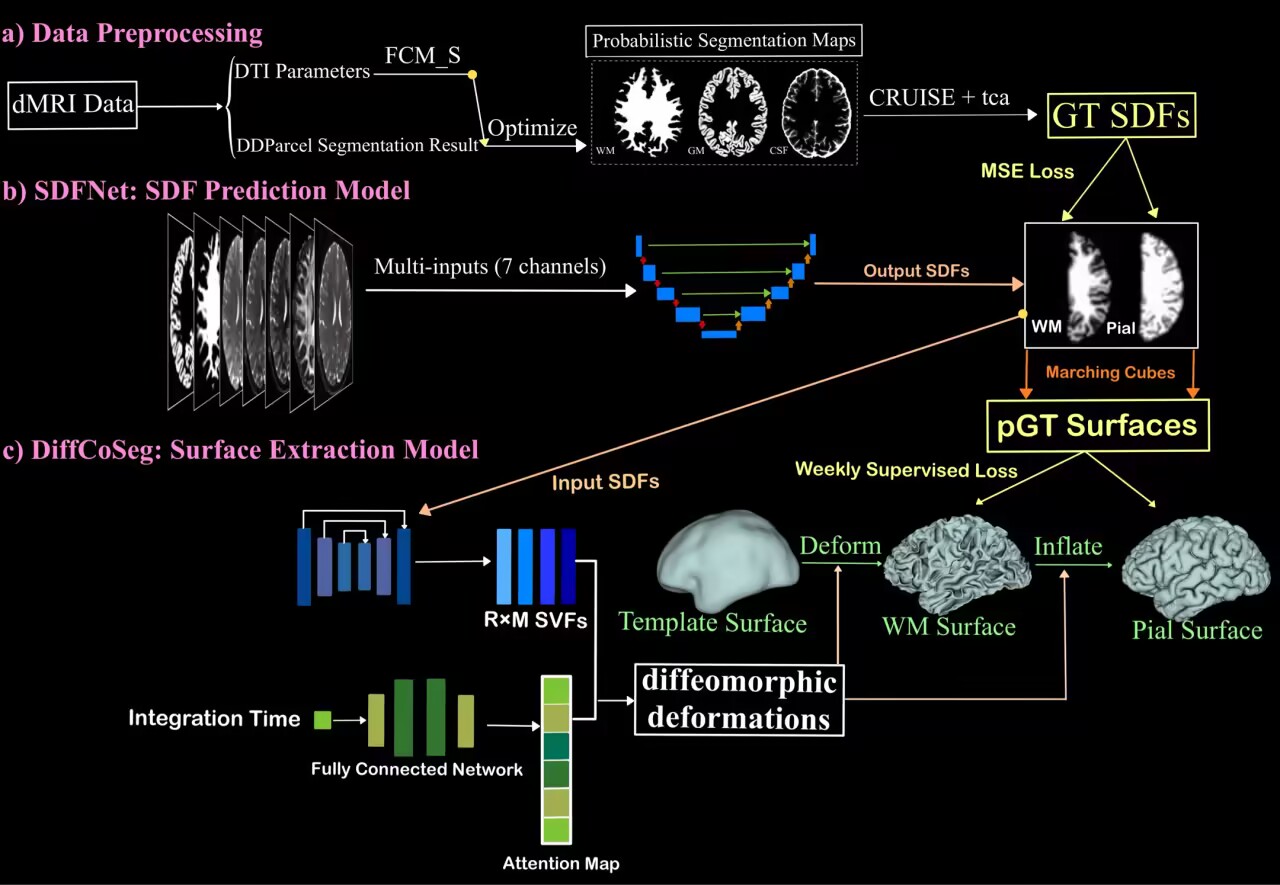}
\caption{Graphic overview of the DDCSR Framework.} \label{fig1}
\end{figure}

DDCSR is a dMRI-based CSR method (see Fig.\ref{fig1}  for an overview), consisting of two major components: 1) an SDF prediction model, \textit{SDFNet}, to compute SDFs for WM and pial surfaces directly from dMRI-derived features, and 2) a surface extraction model, \textit{DiffCoSeg}, to reconstruct the 3D mesh surfaces from the predicted SDFs. 

\subsection{\textbf{Datasets and Preprocessing}}

We utilize the Human Connectome Project Young Adult (HCP-YA) dataset for model training and evaluation \cite{van2013wu}. 150 randomly selected subjects (29.1±3.7 years) are used, of which 60\% are for training, 10\% are for validation, and 30\% are for testing. The acquisition parameters of HCP-YA data are 18 b0 images, 90 b=1000s/mm\textsuperscript{2 }images; TE/TR=89/5520ms; resolution=1.25m\textsuperscript{3}. We perform the following processing steps to generate the input and GT data needed for training the DDCSR model. First, from the dMRI data, we compute five diffusion tensor-based parameter maps (FA, MD, and three eigenvalues). These maps are effective for tissue and brain parcellation \cite{zhang2021deep,zhang2023ddparcel}. Then, from the tensor-based parameter maps, we apply the recently proposed DDParcel \cite{zhang2023ddparcel} to derive the WM and GM probabilistic maps. To improve the delineation of tissue boundaries, the probabilistic maps are processed using fuzzy C-Means with spatial constraints \cite{wen2013brain}. Then, from the processed probabilistic maps, the cortical reconstruction using the implicit surface evolution (CRUISE) method \cite{han2004cruise} is applied to generate SDFs for the WM and pial surfaces for training the SDFNet (Sec \ref{sec:2.2}).

Furthermore, we also use the ABIDE II-NYU dataset \cite{di2017enhancing} and our In-House data scanned at West China Hospital to demonstrate the ability of DDCSR to generalize data from different populations and acquisition protocols. The ABIDE II-NYU dataset consists of 52 subjects with an age range of 10.24 ± 5.74 years. The acquisition parameters include a b-value of 1000 s/mm², 65 directions, and TE/TR = 78/5200 ms, with a resolution of 3 mm³. The In-House data consists of 51 subjects with an age range of 14.00 ± 1.41 years. The acquisition includes a b-value of 1000 s/mm², 64 directions, and TE/TR = 100/9635 ms, with a resolution of 2 mm³.

For experimental evaluation, we also use the T1w data of each testing HCP-YA, ABIDE II-NYU, and In-House dataset. FreeSurfer \cite{fischl2012freesurfer} is performed to perform T1w-based CSR and compute measures including the cortical thickness and surface volumes for experimental evaluation (Sec \ref{sec:3.2}).  
 
 \subsection{\textbf{\textbf{SDFNet: SDFs Prediction} }} \label{sec:2.2}

The SDFNet model (Fig. \ref{fig1}b) is used to predict the SDFs from the dMRI data. Although the above-described CRUISE can generate SDFs from dMRI data, the computational efficiency is inadequate, requiring about 5 minutes to compute SDFs of the WM and pial surfaces. To accelerate this process, we design SDFNet to be an end-to-end prediction network, which requires approximately 3 seconds for SDF generation in the overall pipeline (as shown in Sec \ref{sec:3.2}).

SDFNet takes a multi-channel input, consisting of the b0, FA, MD, the three eigenvalues, and the WM and GM probabilistic maps. The network is based on 3D U-Net \cite{ronneberger2015u} with two modifications. First, to better use the multi-channel input where each channel contains tissue-specific information, the encoder uses a Squeeze-and-Excitation (SE) module \cite{hu2018squeeze}, which dynamically adjusts channel weights via global average pooling and fully connected layers, enhancing the network's ability to focus on important features. Second, to avoid checkerboard artifacts in the output generated using a standard U-Net, the decoder employs an InterpolateConv3D module \cite{mao2019interpolated} that combines trilinear interpolation with convolution to up-sample and extract features. The loss function for training SDFNet is a mean squared error (MSE) loss between the predicted and GT SDFs. In our study, we train two models for the WM and pial SDFs, respectively.

 \subsection{\textbf{\textbf{\textbf{DiffCoSeg: Surface Extraction from dMRI}} }} \label{sec:2.3}

The DiffCoSeg model (Fig. \ref{fig1}c) is used to extract the 3D mesh surfaces from the generated SDFs using SDFNet. DiffCoSeg extends the recent CoSeg method \cite{ma2024weakly} that is designed for T1w-based CSR to process dMRI data. In CoSeg, the input T1w image is processed to learn diffeomorphic deformations using temporal attention networks, which drive the deformation of an initial mesh. A 3D U-Net predicts two stationary velocity fields (SVFs) at each resolution level, with their importance determined by an attention map. pGT cortical surfaces for CoSeg are generated by applying traditional processing methods in FreeSurfer \cite{fischl2012freesurfer} to create ribbon-like structures, which are then converted into pGT surfaces using MC.

 In our DiffCoSeg, we improve the CoSeg model to leverage generated SDFs, which are directly related to the surfaces. First, instead of using T1w images as input, DiffCoSeg uses a dual-channel input consisting of the WM and pial SDFs. Second, for pGT generation, we use predicted SDFs to generate surfaces via MC, instead of using ribbon segmentations. The overall workflow of DiffCoSeg involves two steps: the first step deforms a template mesh to align with the WM surface, while the second step inflates the WM surface to generate the pial surface. The loss of DiffCoSeg incorporates edge length and normal consistency losses \cite{bongratz2022vox2cortex,wickramasinghe2020voxel2mesh} (\(L_{\text{edge}}\), \(L_{\text{nc}}\)) to enhance surface smoothness, along with a bidirectional Chamfer loss for aligning WM surfaces. For pial surfaces, a weakly supervised loss combines boundary and inflation losses (\(L_{\text{inflate}}\) ) \cite{ma2024weakly} to address partial volume effects and accurately capture sulcal details.

 In our DiffCoSeg, we improve the CoSeg model to leverage generated SDFs, which are directly related to the surfaces. First, instead of using T1w images as input, DiffCoSeg uses a dual-channel input consisting of the WM and pial SDFs. Second, for pGT generation, we use generated SDFs to generate surfaces via MC, instead of using ribbon segmentations. The overall workflow of DiffCoSeg involves two steps: the first model deforms a template mesh to align with the WM surface, while the second model inflates the WM surface to generate the pial surface. The loss of DiffCoSeg incorporates edge length and normal consistency losses (\(L_{\text{edge}}\), \(L_{\text{nc}}\)) to enhance surface smoothness, along with a bidirectional Chamfer loss for aligning WM surfaces. For pial surfaces, a weakly supervised loss combines boundary and inflation losses (\(L_{\text{inflate}}\)) to address partial volume effects and accurately capture sulcal details.

 \subsection{\textbf{\textbf{\textbf{Implementation}} }} \label{sec:2.4}

Our method is implemented using Pytorch (v1.7) \cite{paszke2019pytorch}, and all computation is performed on a Linux workstation equipped with NVIDIA RTX 3090 GPUs. For both SDFNet and DiffCoSeg, Adam is used as the optimizer, with a total of 400 epochs, a learning rate of 0.0001, and a batch size of 1. For the SDFNet model, we use Advanced Normalization Tools (ANTs) \cite{avants2009advanced} to
rigidly align the baseline dMRI images to the MNI template for preprocessing.
The transformation matrices derived during this alignment are applied to all input
images. For the DiffCoSeg model, the SDFs inputted for training are also in MNI space. The weights for \(L_{\text{edge}}\), \(L_{\text{nc}}\), and \(L_{\text{inflate}}\) are chosen as 0.5, 2.5, and 5.0, respectively. An initial 160k MNI template vertex mesh \cite{ma2024weakly} is used. The pial surface model undergoes 30 epochs of MSE pre-training before being further trained with weakly supervised loss.

 \section{\textbf{Experiments} }
 \subsection{\textbf{\textbf{Experimental Design} }} \label{sec:3.1}

We compare DDCSR with the following three methods. First, as a baseline method, we include the widely used T1w-based method that performs CSR on T1w using FreeSurfer and registers to the dMRI space. Second, we include a dMRI-based CSR method using traditional step-by-step processing that computes SDFs using CRUISE \cite{han2004cruise}, followed by surface extraction using MC \cite{lorensen1998marching}. Third, we include a dMRI-based CSR method using a deep CoSeg network to directly generate the cortical surface from dMRI-derived parameters.

We first perform the experimental comparison on the HCP-YA data. Due to the high-quality acquisition protocol and well-established processing pipeline used for the HCP-YA data, the registration between the dMRI and T1w images is highly accurate. Therefore, we choose the T1w-based approach as a reference to evaluate the other three methods (i.e., CRUISE+MC, CoSeg, DDCSR). Five evaluation metrics are computed between the T1w- and dMRI-based surfaces, including Chamfer Distance (CH), Hausdorff Distance (HD), Dice Similarity Coefficient (DSC), Volume Similarity (VS), and Thickness Similarity (TS). Specifically, CH and HD measure the geometric similarity, DSC measures the spatial overlap, and VS and TS measure the similarity of the cortical morphometry. Note that CH, HD, and DSC are computed between the coregistered dMRI and T1w surfaces, using the T1w-based surfaces as ground truth. VS and TS do not rely on the registration and are computed from the T1w and dMRI surfaces separately.

We then assess the generalization ability of the compared methods on the ABIDE and In-House datasets. Since these two datasets are acquired using clinically standard protocols, the registration between T1w and dMRI is suboptimal. Therefore, we only include the quantitative measures against the T1w that do not rely on registration, i.e., VS and TS.

 \subsection{\textbf{\textbf{\textbf{Results on HCP-YA data}} }} \label{sec:3.2}

\begin{figure}[H]
 \centering
\includegraphics[width=0.7\textwidth]{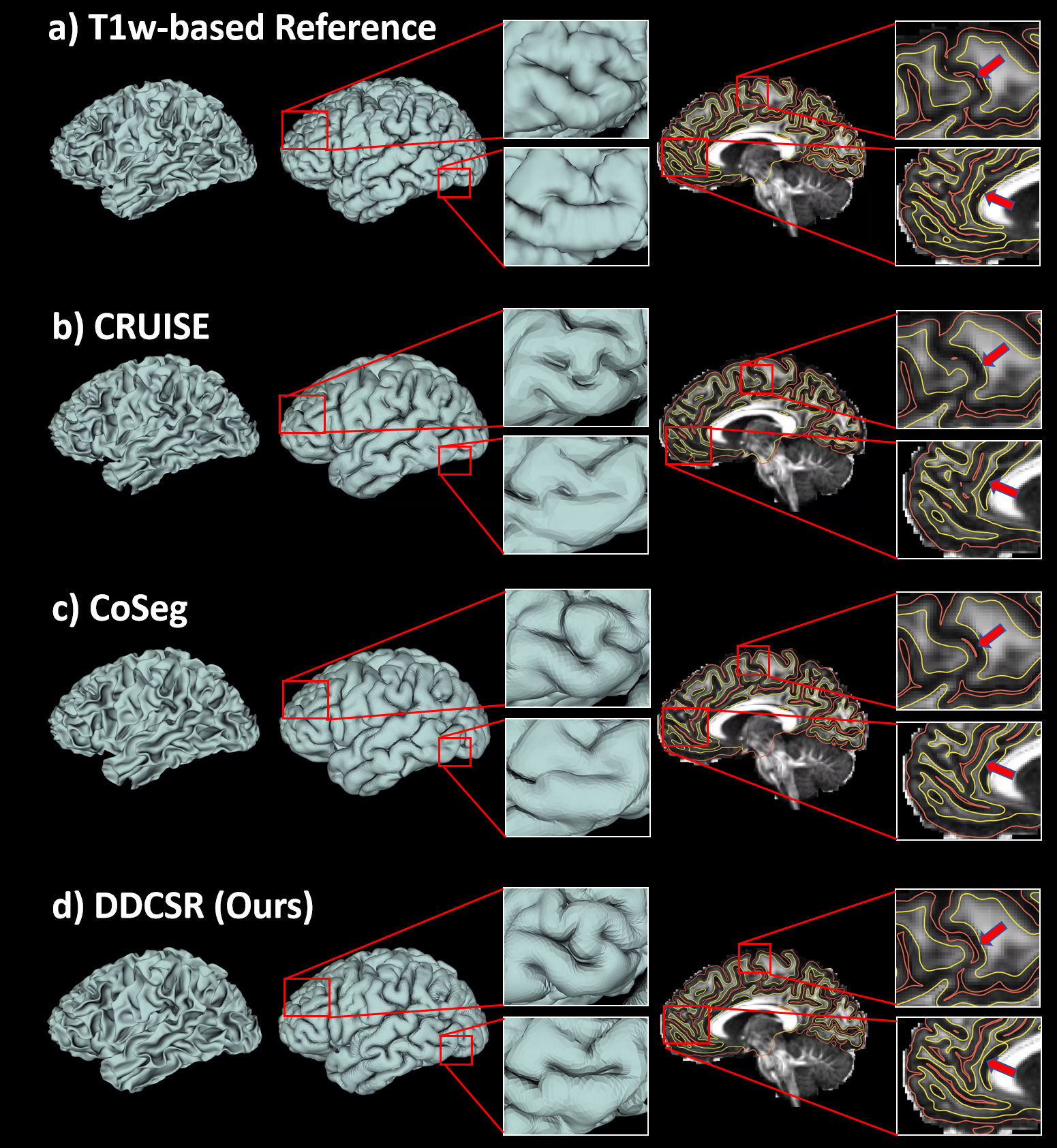}
\caption{Surface reconstruction comparison across the different methods. From left to right: WM surface, pial surface, slice view (yellow contour: WM surface; red contour: pial surface). The background is based on FA.} \label{fig2}
\end{figure}

Table 1 gives the quantitative results across the compared methods of the HCP-YA dataset. DDCSR outperforms both CRUISE and CoSeg across all five evaluation metrics. Fig \ref{fig2} gives the 3D mesh and 2D slice visualization of the reconstructed WM and pial surfaces from each method. We can observe that even though the reference surface is generated from high-quality MRI data but has misalignments in local regions. CRUISE and CoSeg can approximate the surfaces well but struggle with sulcal details. Overall, the results demonstrate that DDCSR excels in capturing complex cortical folds, and shows a high degree of alignment with the underlying anatomical structures.

\begin{table}[ht]
\caption{Comparison of surface reconstruction methods on the WM and pial surfaces. ↓ indicates that lower value is better and ↑ indicates that higher value is better.}
\label{tab:comparison}
\centering
\begin{tabular}{|l|c|c|c|c|c|c|}
\hline
\multicolumn{5}{|c|}{WM Surface} & \multirow{2}{*}{Runtime (s) $\downarrow$} \\ \cline{1-5} 
Method & CH (mm) $\downarrow$ & HD (mm) $\downarrow$ & DSC $\uparrow$ & VS $\uparrow$ & \\ \hline
CRUISE & 1.756 $\pm$ 0.168 & 8.369 $\pm$ 1.066 & 0.78 $\pm$ 0.02 & 0.906 $\pm$ 0.045 & 267.20 $\pm$ 9.45 \\ 
CoSeg & 1.438 $\pm$ 0.826 & 7.379 $\pm$ 1.113 & 0.77 $\pm$ 0.02 & 0.960 $\pm$ 0.021 & \textbf{3.17 $\pm$ 0.39} \\ 
\textbf{DDCSR} & \textbf{1.432 $\pm$ 0.425} & \textbf{6.823 $\pm$ 1.631} & \textbf{0.82 $\pm$ 0.07} & \textbf{0.979 $\pm$ 0.008} & 5.38 $\pm$ 0.34 \\ \hline
\multicolumn{5}{|c|}{Pial Surface} & \multirow{2}{*}{TS $\uparrow$} \\ \cline{1-5} 
Method & CH (mm) $\downarrow$ & HD (mm) $\downarrow$ & DSC $\uparrow$ & VS $\uparrow$ & \\ \hline
CRUISE & 2.453 $\pm$ 0.181 & 9.542 $\pm$ 1.195 & 0.74 $\pm$ 0.04 & 0.955 $\pm$ 0.021 & 0.5250 $\pm$ 0.036 \\ 
CoSeg & 1.804 $\pm$ 1.146 & 8.259 $\pm$ 1.208 & 0.79 $\pm$ 0.05 & 0.968 $\pm$ 0.009 & 0.916 $\pm$ 0.020 \\ 
\textbf{DDCSR} & \textbf{1.757 $\pm$ 0.653} & \textbf{7.916 $\pm$ 1.122} & \textbf{0.81 $\pm$ 0.08} & \textbf{0.987 $\pm$ 0.019} & \textbf{0.939 $\pm$ 0.034} \\ \hline
\end{tabular}
\end{table}\label{table1}

 \subsection{\textbf{\textbf{\textbf{Results on ABIDE and In-House Data}} }} \label{sec:3.3}

\begin{figure}[H]
  \centering
\includegraphics[width=\textwidth]{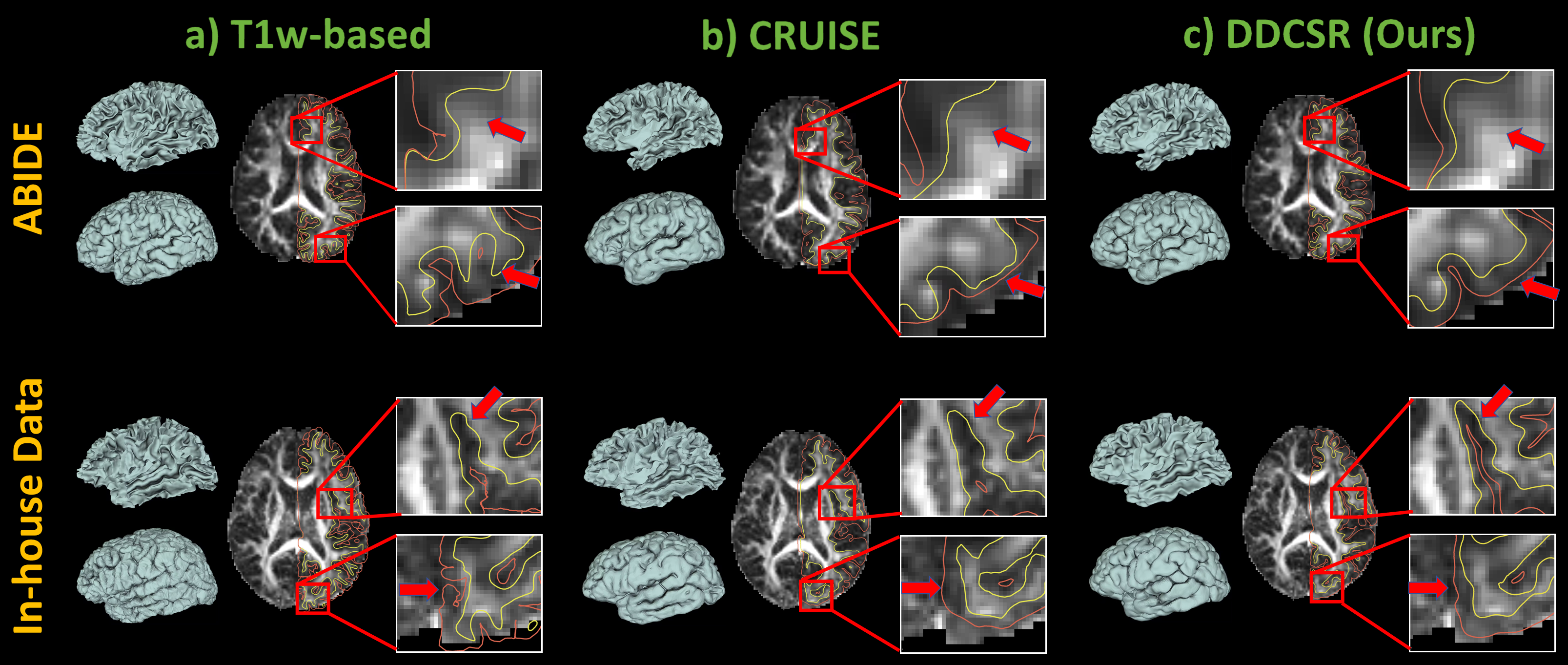}
\caption{Surface reconstruction on example ABIDE and In-House datasets. From left to right: WM surface (up), pial surface (down), slice view (yellow contour: WM surface; red contour: pial surface). The background is based on FA.} \label{fig3}
\end{figure}

We evaluate our method on the ABIDE dataset and our In-House data, which are acquired using different protocols from that of the HCP dataset. The results in Fig \ref{fig3} show that the reference surfaces generated through inter-modality registration using ANTs exhibit several misalignments. The surfaces produced using the CRUISE lacked precision in contour definition and sulcal details. When applying the HCP-trained CoSeg model directly to the ABIDE and In-House data, it results in distorted surfaces (thus not included in the figure). In contrast, the DDCSR method performs well, maintaining accurate surface reconstruction and effectively capturing sulcal details. The quantitative measures TS and VS for CRUISE and DDCSR compared to the reference surfaces are as follows: \textbf{CRUISE}: TS = 0.467 ± 0.024, VS (WM) = 0.894 ± 0.029, VS (pial) = 0.921 ± 0.027; \textbf{DDCSR}: TS = 0.869 ± 0.096, VS (WM) = 0.966 ± 0.016, VS (pial) = 0.952 ± 0.024. Overall, our DDCSR largely outperforms the compared CRUISE and CoSeg methods.

 \section{\textbf{Conclusion} }

This paper introduces DDCSR, a deep learning framework for CSR from dMRI data. Unlike traditional methods, DDCSR eliminates the need for anatomical T1w data by directly working with dMRI data. Our experimental comparison shows that DDCSR achieves accurate CSR and excels at capturing fine details of sulcal and gyral patterns. Additionally, DDCSR exhibits strong generalization capabilities across diverse populations and datasets. Finally, DDCSR enables fast and efficient processing, making it a promising tool for large-scale neuroimaging studies. Future work could include investigation of testing on brains with lesions (such as tumors or edema) and/or from different age ranges (e.g., babies).

\section*{Acknowledgments}
This work is supported in part by National Key R\&D Program of China (No. 2023YFE0118600), National Natural Science Foundation of China (No. 62371107), National Institutes of Health (R01MH108574, P41EB015902, R01MH074794, R01MH125860, R01MH119222, R01MH132610, R01NS125781).

\bibliographystyle{unsrt}  
\bibliography{references}

\end{document}